\documentclass[aps,prb,twocolumn,superscriptaddress,showpacs,floatfix]{revtex4}
\usepackage{graphicx}
\usepackage{epsfig}
\usepackage{amsmath}
\usepackage{amssymb}
\usepackage{subfigure}

\begin{document}

\frenchspacing

\title{Pressure-induced softening as a common feature of framework structures that have negative thermal expansion}

\author{Hong Fang}
\affiliation{Department of Earth Sciences, University of Cambridge, Downing Street, Cambridge CB2 3EQ, U.K.}

\author{Martin T. Dove}
\email{martin.dove@qmul.ac.uk}
\affiliation{Department of Earth Sciences, University of Cambridge, Downing Street, Cambridge CB2 3EQ, U.K.}
\affiliation{Centre for Condensed Matter and Materials Physics, School of Physics and Astronomy, Queen Mary University of London, Mile End Road, London E1 4NS, U.K.}

\date{\today}
\begin{abstract}
Results of a series of molecular dynamics simulations of cubic siliceous zeolites suggest that pressure-induced softening -- the phenomenon in which a material becomes progressively more compressible under pressure -- is likely to be a common feature of framework materials that show negative thermal expansion. The correlation between the negative thermal expansion and the pressure-induced softening is investigated on the basis of thermodynamics.
\end{abstract}

\pacs{62.20.-x, 65.40.De, 64.30.-t}

\maketitle

\section{Introduction}

Almost all materials become stiffer when compressed, as a result of the constituent atoms being squeezed together. It therefore comes as something of a shock that some materials -- among them amorphous silica \cite{Tsiok 1998}, ZrW$_2$O$_8$  \cite{Pantea 2006} and Zn(CN)$_2$ \cite{Chapman 2007} -- actually become \textit{softer} under compression. Formally, the stiffness is defined through the zero-pressure bulk modulus $B_0 = V_0 (\partial V_0 / \partial P)_T^{-1}$, and the change in stiffness is defined through the differential $B_0^\prime = \partial B_0/\partial P$, which is a positive quantity for almost all materials. However, in these cited examples it is found that $B_0^\prime$ has a negative value. There is as yet no theoretical explanation for this effect, which can be called ``pressure-induced softening'', but in a simulation study of pressure-induced softening in amorphous silica \cite{Walker 2007} we drew attention to the role of fluctuations involving whole-body rotations of SiO$_4$ tetrahedra. Given that the same fluctuations are implicated in the similarly counter-intuitive phenomenon of negative thermal expansion (NTE) \cite{Lind 2012}, and given that the few materials in which pressure-induced softening has been identified also show NTE, we suggest that most NTE materials will show pressure-induced softening, and in this paper we demonstrate the plausibility of this hypothesis.

The basis for linking pressure-induced softening with negative thermal expansion can be understood by considering the relatively simple example of Zn(CN)$_2$.\cite{Goodwin 2005} Its perfect structure has linear Zn--C--N--Zn linkages of bonds along the crystallographic $\left< 1,1,1 \right>$ directions. Uniform compression of the perfect structure will force compression of these bonds, which are stiff and will become stiffer on further compression. Hence at a temperature of 0 K, or in a static lattice energy calculation, we might expect to find a positive value of $B_0^\prime$. However, on heating thermal fluctuations will cause instantaneous buckling of the Zn--C--N--Zn linkages -- a process aided by the fact that the rigid-unit-mode flexibility of the structure allows for localised distortions \cite{Goodwin2006} -- so that an external compressive force can be accommodated with relatively low energy cost by further buckling without the need to compress the individual bonds. If we now consider the case of stretching the structure (application of negative pressure), the stretch will first be accommodated by reducing the buckling of the linkages of bonds, but when the buckling has been stretched out the second process is to stretch the individual bonds. This will cost a lot more energy, and the volume change per unit of stretch force will reduce. This means that the bulk modulus will increase on stretching, and hence we have a negative value of $B_0^\prime$.

Because the fluctuations that buckle the linkages give rise to a reduction in crystal volume in many framework, and because their amplitude increases with temperature, we have the possibility -- perhaps in some cases inevitability -- for negative thermal expansion. Thus we might expect pressure-induced softening to be linked to NTE. Put another way, we might expect that many NTE materials will also show pressure-induced softening. In the case of Zn(CN)$_2$, experimentally it is found to have large NTE, and a negative value of $B_0^\prime$, i.e.\ pressure-induced softening.\cite{Chapman 2007} Our recent simulation study of Zn(CN)$_2$ \cite{our_MD} is consistent with the experimental data \cite{our_Exp} and shows that $B_0^\prime$ has a dependence on temperature of the form described above.

In this paper we test the proposal of a direct link between NTE and pressure-induced softening by performing simulation experiments on the full suite of zeolites with cubic lattice symmetry \cite{zeolite_note}. Zeolites are low-density framework structures built from corner-linked SiO$_4$ tetrahedra, many of which are found naturally with ionic substitution on the tetrahedral site (e.g.\ Al for Si) with associated charge-balancing cations (e.g.\ Na) found in the large pores in the structure. It has long been recognised that some zeolites show NTE \cite{Miller 2009,Lightfoot 2001}, although there has not yet been a systematic study of the set of the cubic zeolites. Here we make two predictions, first that most cubic zeolites will show NTE, and second, based on the preceding discussion, that those that do have NTE will also show pressure-induced softening. This is much easier tested by molecular dynamics simulation than experiment, and for siliceous zeolites we have some good force fields derived from quantum mechanical calculations and tested in many independent studies. There are 13 candidate zeolites with crystal structures of cubic symmetry and fully connected SiO$_4$ tetrahedra, all of which are investigated in the current work. For reference for the rest of this paper, we note that all zeolites are assigned a three-letter name\cite{zeolite_note}, sometimes which relates to a historical name (e.g.\ ANA for analcime, FAU for faujisite).

The thermodynamic theory to link NTE with pressure-induced softening is derived in Section II. Section III gives the computational method and Section IV presents the main results. Conclusions are drawn in Section V.

\section{Thermodynamic Background}

The Helmholtz free energy of an insulating crystal \cite{Ashcroft 1976} in the classical high-temperature approximation is written as

\begin{eqnarray}\label{eq4}
F = \Phi  + \sum\limits_s {\ln \left( {\frac{{\hbar \omega _s }}{\tau }} \right)},
\end{eqnarray}

\noindent where the first term on the right-hand side is the lattice energy of the crystal at zero temperature. The second term involves the sum over all wave vectors $\textbf{k}$ on all branches of the phonon dispersion curves $j$, with $\omega_s$ as the frequency of each phonon mode denoted by $s=\{j,\textbf{k}\}$. $\tau=k_\mathrm{B}T$ is the temperature in the unit of energy.

At equilibrium, the pressure $p$ is obtained as the derivative of the free energy with respect to the crystal volume $V$:

\begin{eqnarray}\label{eq6}
p = - \frac{{\partial F}}{{\partial V}}= - \frac{{\partial \Phi }}{{\partial V}} + \frac{3N\tau }{V}\overline \gamma,
\end{eqnarray}

\noindent The overall Gr{\"u}neisen parameter $\overline \gamma$ is defined as the sum over all the mode Gr{\"u}neisen parameters $\gamma_s=- \left( {V/\omega _s } \right)\left( {\partial \omega _s /\partial V} \right)$:

\begin{eqnarray}\label{eq7}
\overline \gamma   = \sum\limits_s {\gamma _s } /\left( {3N} \right)
\end{eqnarray}

\noindent with $N$ as the total number of atoms in the system. The bulk modulus of the material can be calculated using

\begin{eqnarray}\label{eq11}
 B =  - V\frac{{\partial p}}{{\partial V}} = V\frac{{\partial ^2 \Phi }}{{\partial V^2 }} + \frac{3N\tau }{V}\overline\gamma  + \frac{{3NB\tau }}{V}\frac{{\partial \overline \gamma }}{{\partial p}},
\end{eqnarray}

\noindent where we have used

\begin{eqnarray}\label{eq12}
\frac{{\partial \overline \gamma }}{{\partial V}} = \frac{{\partial \overline \gamma }}{{\partial p}}\frac{{\partial p}}{{\partial V}} =  - \frac{B}{V}\frac{{\partial \overline \gamma }}{{\partial p}}.
\end{eqnarray}

\noindent Thus,

\begin{eqnarray}\label{eq13}
B = \frac{{V\partial ^2 \Phi /\partial V^2  + \left( {3N\tau /V} \right)\overline\gamma }}{{1 - \left( {3N\tau /V} \right)\left( {\partial \overline \gamma /\partial p} \right) }}.
\end{eqnarray}

\noindent From Equation~\ref{eq13}, we can obtain the first derivative of the bulk modulus with respect to pressure as

\begin{widetext}
\begin{eqnarray}\label{eq14}
B^{\prime} = \frac{{\partial B}}{{\partial p}} = \frac{{\partial B}}{{\partial V}}\frac{{\partial V}}{{\partial p}} \approx  - \frac{V}{B}\left( {\frac{{\partial ^2 \Phi }}{{\partial V^2 }} + \frac{{V\partial ^3 \Phi }}{{\partial V^3 }}} \right) + \frac{{3N\tau }}{{V }}\frac{{\overline \gamma  }}{B} + \frac{{6N\tau }}{{V }}\frac{{\partial \overline \gamma  }}{{\partial p}} + \frac{{3N\tau B}}{{V}}\frac{{\partial ^2 \overline \gamma  }}{{\partial p^2 }}.
\end{eqnarray}
\end{widetext}

\noindent In Equation~\ref{eq14}, we have used the approximation

\begin{eqnarray}\label{eq16}
\left| {\frac{{3N\tau }}{{V }}\frac{{\partial \overline \gamma  }}{{\partial p}}} \right| \ll 1
\end{eqnarray}

\noindent which is generally valid for the zeolites we have studied here, as will be seen in Table~\ref{table1} in Section V. According to this, from Equation~\ref{eq13}, one has

\begin{eqnarray}\label{eq17}
B \approx V\frac{{\partial ^2 \Phi }}{{\partial V^2 }} + \frac{3N\tau }{V}\overline\gamma.
\end{eqnarray}

\noindent Accordingly, we can rewrite Equation~\ref{eq14} in the more compact form

\begin{eqnarray}\label{eq18}
B^{\prime} = B^{\prime}\left.\right|_{T = 0}  + \frac{{3N\tau }}{{V}}\left[ {\frac{{\overline \gamma  }}{B} + 2\frac{{\partial \overline \gamma  }}{{\partial p}} + B\frac{{\partial ^2 \overline \gamma  }}{{\partial p^2 }}} \right],
\end{eqnarray}

\noindent where the first term on the right-hand side, namely

\begin{eqnarray}\label{eq19}
B^{\prime}\left.\right|_{T = 0}  =  - \frac{V}{B}\left( {\frac{{\partial ^2 \Phi }}{{\partial V^2 }} + \frac{{V\partial ^3 \Phi }}{{\partial V^3 }}} \right)
\end{eqnarray}

\noindent is the value of $B^{\prime}$ at zero temperature obtained in a harmonic-lattice-dynamics calculation. In our case, we will calculate this term for all the cubic NTE zeolites in harmonic lattice dynamics using a force field with the Buckingham potential.

Thus, according to Equation~\ref{eq18}, if $B^{\prime}\left.\right|_{T = 0}$ is positive, given that all the rest terms are negative, $B^{\prime}$ may become negative when the temperature is high enough. In fact, since NTE materials have negative coefficient of thermal expansion, one should have

\begin{eqnarray}\label{eq20}
\overline \gamma   < 0
\end{eqnarray}

\noindent and we will see in Section V, that for all the cubic NTE zeolites,

\begin{eqnarray}\label{eq21}
 \frac{{\partial \overline \gamma  }}{{\partial p}} < 0 \nonumber \\
 \frac{{\partial ^2 \overline \gamma  }}{{\partial p^2 }} < 0 .
\end{eqnarray}

\noindent Note that, if $\overline \gamma$, $\partial \overline \gamma  /\partial p$ and $\partial ^2 \overline \gamma  /\partial p^2$ are in the same order of magnitude, the term containing $\partial ^2 \overline \gamma  /\partial p^2$ will contribute dominantly to a negative $B^{\prime}$ due to its large coefficient involving the bulk modulus $B$.

\begin{figure}[t]
\begin{center}
\includegraphics[width=8.0cm]{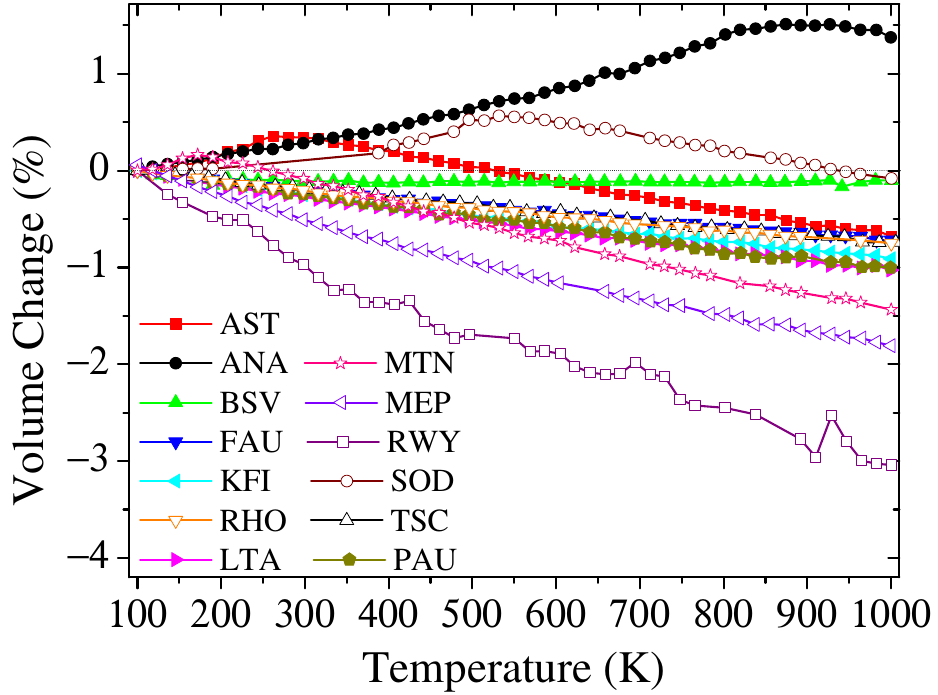}
\end{center}
\caption{\label{fig:VvsT_plots} Calculated V-T plots from 100 to 1000 K at ambient pressure for all the cubic zeolites. ANA shows positive thermal expansion. BSV shows almost zero thermal expansion. AST, MTN and SOD show negative thermal expansion only in their high-temperature phases.}
\end{figure}

\section{Computational Methods}

The molecular dynamics (MD) simulations were carried out using DL$\_$POLY \cite{Todorov 2006}. The O--O, Si--O and Si--Si interactions are described by Coulomb interactions and Buckingham potentials $\phi(r)$ of the form

\begin{equation}
\phi_{ij}(r_{ij}) = A_{ij} \exp(-r_{ij}/\rho_{ij}) - C_{ij}r^{-6}_{ij} ,
\end{equation}

\noindent where $r_{ij}$ is the distance between two atoms of type $i$ and $j$, and the parameters $A_{ij}$, $\rho_{ij}$ and $C_{ij}$ for each atom pair type are taken from the force field of Tsuneyuki\cite{Tsuneyuki 1988}. The point charges on theSi and O atoms are (in the units of electron charge) $2.4$ and $-1.2$ respectively. The long-range Coulomb energy was calculated using the Ewald method with precision of 10$^{-4}$. Typical simulations, lasting around 30 ps in the production stage with a 10 ps equilibration stage, were performed using time steps of $0.001$ ps and the velocity Verlet scheme \cite{William 1982}. Long-time stability of the structures at high temperature were tested up to 200 ps. Simulations were performed using the Nos\'{e}--Hoover constant-pressure constant-temperature ensemble \cite{Hoover 1985}, with relaxation times of $1.0$ ps for both thermostat and barostat. The first suite of simulations were performed at constant pressure to search for NTE, followed by a large number of simulations over a range of pressure at a fixed temperature for a large number of different temperature values. Typical sample sizes were $6\times6\times6$ unit cells.

The calculations of the density of states and the Gr\"{u}neisen parameters for the studied zeolites were carried out in harmonic lattice dynamics using GULP \cite{Gale 1997}. The same potential model as in the MD was used in these calculations.

\begin{figure}[t]
\begin{center}
\includegraphics[width=8.0cm]{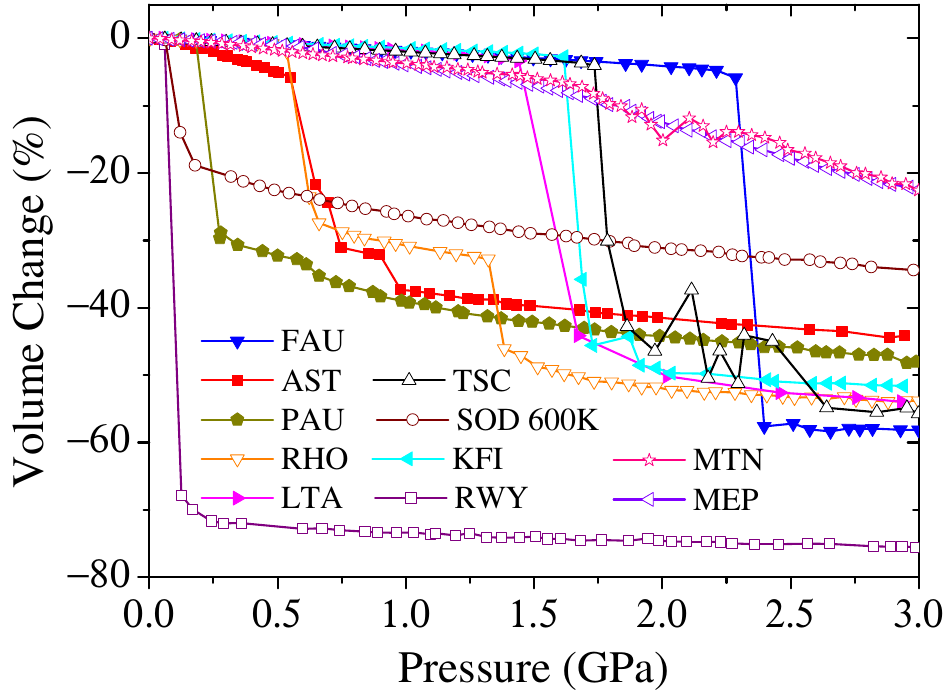}
\end{center}
\caption{\label{fig:VvsP_plots} Calculated V-P plots from 0.0 to 3.0 GPa at 300 K for the 11 NTE zeolites. RWY, SOD and PAU undergo phase transitions at very low pressure ($< 0.2$ GPa).}
\end{figure}

\section{Results}

\subsection{Search for negative thermal expansion in cubic zeolites}

Figure~\ref{fig:VvsT_plots} shows the simulated volume--temperature relationships at ambient pressure for all cubic zeolites. Only ANA has positive thermal expansion throughout the temperature range. BSV shows almost zero thermal expansion. The other 11 zeolites have NTE. Three of these undergo phase transitions, and it is their high-temperature phases that show NTE. Figure~\ref{fig:VvsP_plots} shows the simulated volume--pressure relationships at a temperature of 300~K. Inevitably most examples show a pressure-induced phase transition. RWY, SOD and PAU undergo phase transitions at very low pressure ($< 0.2$ GPa). Thus, from the original pool of 13 zeolites we have 8 useful candidate materials where the transition pressure is not too low and in which we can explore the link between pressure-induced softening and NTE. The key results from these preliminary temperature and pressure scans are summarised in Table~\ref{results_table}.

\begin{figure}[t]
\begin{center}
\includegraphics[width=8.0cm]{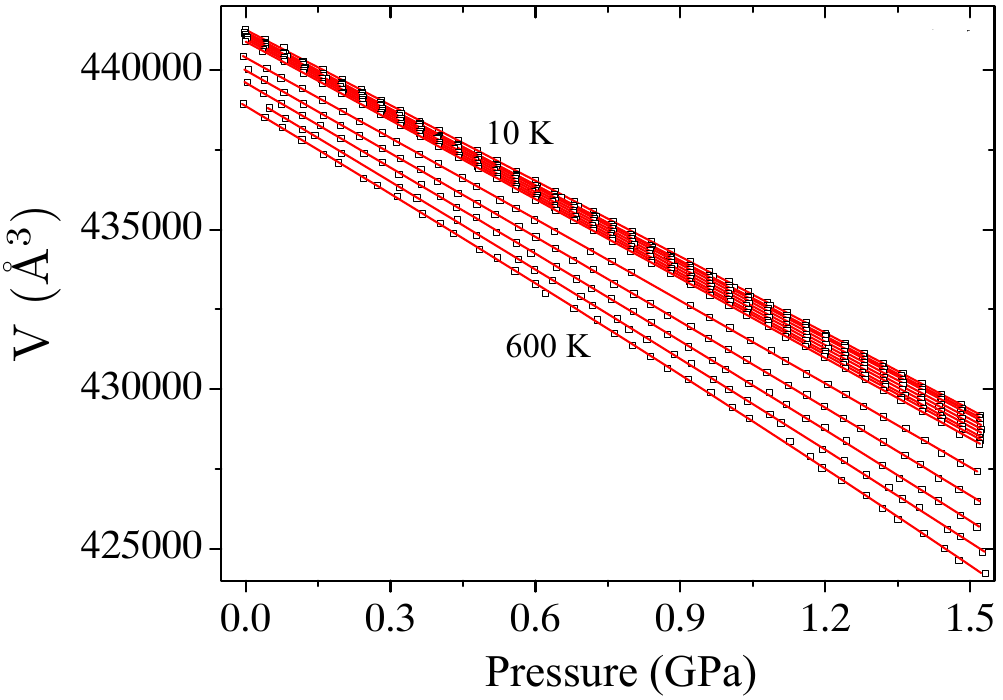}
\end{center}
\caption{\label{fig:FAU_isotherms} Fitted isotherms of FAU at different temperatures from 1 to 600 K with a 10 K interval from 10 to 100 K and an 100 K increment from 100 to 600 K, using the 3rd-order BM EoS. The convex-parabola trend of all the isotherms indicate negative $B^{\prime}_0$ and pressure-induced softening of the material. The value of volume in the plot is for the supercell used in MD.}
\end{figure}

We found that almost all the modes with negative Gr\"{u}neisen parameters are rigid unit modes (RUMs) in these cubic zeolites. Calculations using the CRUSH code \cite{Giddy 1993,Hammonds 1994} reveal that these RUMs correspond to the rotations and the translations of the SiO$_4$ tetrahedra, with the translational modes occupy the lowest energy band. The big difference between the positive-thermal-expansion ANA and other NTE cubic zeolites is that the low-frequency translational RUMs of ANA do not have large negative Gr\"{u}neisen parameters, while the medium and high-frequency non-RUMs in ANA have relatively large positive Gr\"{u}neisen parameters. The low-frequency translational RUMs contribute the most to the NTE of the material, which is similar to the finding in our work on Zn(CN)$_2$ \cite{our_MD}. Detailed discussions of the origins of NTE of these cubic zeolites will be given elsewhere \cite{prepare}.

\subsection{Pressure-induced softening in NTE-zeolites}

From sequences of pressure-sweep simulations over a range of fixed temperatures we have obtained values of $V_0$, $B_0$ and $B_0^\prime$ by fitting isothermal data using the equation of states (EoS). We have explored several EoS formalisms \cite{EoSreview}, including 3rd and 4th-order Birch-Murnaghan (BM) EoS, Vinet (Universal) EoS, and Keane EoS. It was found that the 3rd and 4th-order BM EoSs have the greater stability. The Vinet and Keans EoSs gave similar results to those of the BM EoS. The 3rd-order BM gave excellent results for most of the studied zeolites, but significant improvements were obtained using the 4th-order BM for AST and MTN \cite{SI}. The fitted values of $B_0$ and $B_0^\prime$ at 300 K are given in Table~\ref{results_table}.

\begin{table}[t]
\setlength{\tabcolsep}{3pt}
\caption{Calculated coefficients of thermal expansion $\alpha_\mathrm{V}$, phase-transition temperature $T_\mathrm{c}$ at ambient pressure, and phase-transition pressure $P_\mathrm{c}$ at 300 K. Because $\alpha_\mathrm{V}$ varies with temperature, the averaged values over the temperature range are given. For AST and MTN which only have NTE in their high temperature phases, the values are averaged over 300-1025 K. The bulk modulus at 0 pressure $B_0$ and its first derivative with respect to pressure $B_0^\prime$ of each NTE zeolite is obtained from fitting the 300 K isotherm with the 3rd-order BM EoS. These values are not given for RWY, SOD and PAU because they quickly go through phase transitions on compression.}
\centering
\begin{tabular}{c| c| c| c| c| c}
  \hline\hline
  Zeolite & $\alpha_\mathrm{V}$ (MK$^{-1}$) & $T_\mathrm{c}$ (K) & $P_\mathrm{c}$ (GPa)  & $B_0$ & $B_0^\prime$ \\ [0.5ex]
  \hline
  AST & $-13.8$ & 250 & 0.5 & 16(2) & $-28(9)$\footnotemark[6] \\ [0.5ex]
  FAU & $-8.0$\footnotemark[1] & -- & 2.2  & 51.1(2)\footnotemark[4] & $-2.6(3)$ \\ [0.5ex]
  KFI& $-10.0$ & -- & 1.6 & 65.7(5) & $-4.4(9)$  \\ [0.5ex]
  RHO& $-8.3$\footnotemark[2] & -- & 0.5 & 59(1) & $-4(4)$ \\ [0.5ex]
  LTA& $-11.3$  & -- & 1.5 & 61.8(6)\footnotemark[5] & $-13.0(9)$  \\ [0.5ex]
  MTN& $-19.3$\footnotemark[3]  & 200 & 1.5  & 27(1) & $-3(2)$\footnotemark[6] \\ [0.5ex]
  MEP& $-20.8$  & -- & 0.6  & 58(2) & $-30(7)$ \\ [0.5ex]
  TSC& $-8.2$   & -- & 1.75 & $52.4(2)$ & $-4.9(2)$ \\ [0.5ex]
  RWY& $-35.2$   & -- & $< 0.1$  & -- & -- \\ [0.5ex]
  SOD& $13.4$   & 600 & $< 0.1$  & -- & -- \\ [0.5ex]
  PAU& $-11.2$  & -- & $< 0.2$ & -- & --  \\ [0.5ex]
  BSV& $-0.9$   & -- & --  & 56.2(4) & $-0.03(37)$ \\ [0.5ex]
  ANA& 19.7    & 800 & 1.2 & 10.3(3) & 0.3(3)  \\ [0.5ex]
  \hline
\end{tabular}
\footnotetext[1]{Experimental value \cite{Attfield 1998} averaged over 25-573 K is $-12.6$ MK$^{-1}$.}
\footnotetext[2]{Averaged value \cite{Tschaufeser 1995} over 0-500 K calculated from lattice dynamics in quasi-harmonic approximation is $-13$ MK$^{-1}$.}
\footnotetext[3]{Experiment shows that MTN goes through a phase transition at about 370 K and only its high temperature phase shows NTE. The experimental value \cite{Park 1997} averaged over 463-1002 K is $-5.0$(0.7) MK$^{-1}$.}
\footnotetext[4]{Experimental value is $38$(2) GPa \cite{Colligan 2004}.}
\footnotetext[5]{DFT calculated value is $46$ GPa \cite{Astala 2004}.}
\footnotetext[6]{The fitted values of $B_0^\prime$ of AST and MTN using the 4th-order BM EoS are $-$108(7) and $-$37(4), respectively \cite{SI}.}
\label{results_table}
\end{table}

\begin{figure}[t]
\begin{center}
\includegraphics[width=8.0cm]{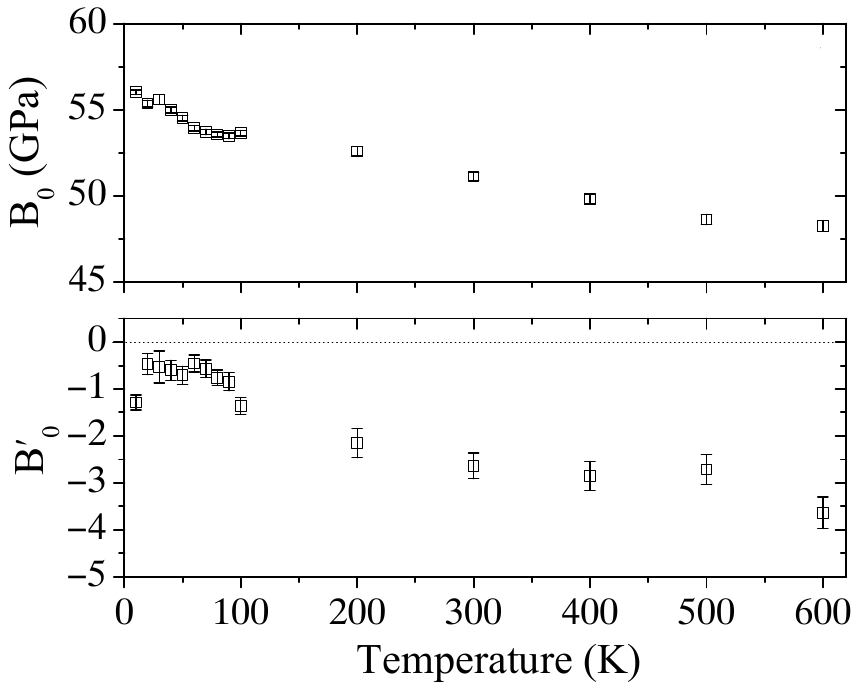}
\end{center}
\caption{\label{fig:FAU_results} Fitted $B_0$ and $B^{\prime}_0$ of FAU changing with temperature. Obviously, the average value of $B^{\prime}_0$ throughout the temperature is negative, indicating pressure-induced softening of the material.}
\end{figure}

As an example, Figure~\ref{fig:FAU_isotherms} shows the fitting to isotherms of FAU at different temperatures using the 3rd-order BM EoS. Figure~\ref{fig:FAU_results} shows the temperature-dependence of the fitted $B_0$ and $B^{\prime}_0$ values. In this case, the fitted value of $B^{\prime}_0$ is negative at all temperatures; in Figure~\ref{fig:FAU_isotherms} a convex-parabola trend seen in the isotherms shows that the volume contracts more rapidly at higher pressure, and this gives the negative value of $B^{\prime}_0$. In this regard, a 3rd-order BM EoS was adequate to describe this trend in the isotherm. Plots of the fitted isotherms, equilibrium volume $V_0$, $B_0$ and $B^{\prime}_0$ for all the systems are given in the supplemental material \cite{SI}.

\begin{figure}[t]
\begin{center}
\includegraphics[width=8.5cm]{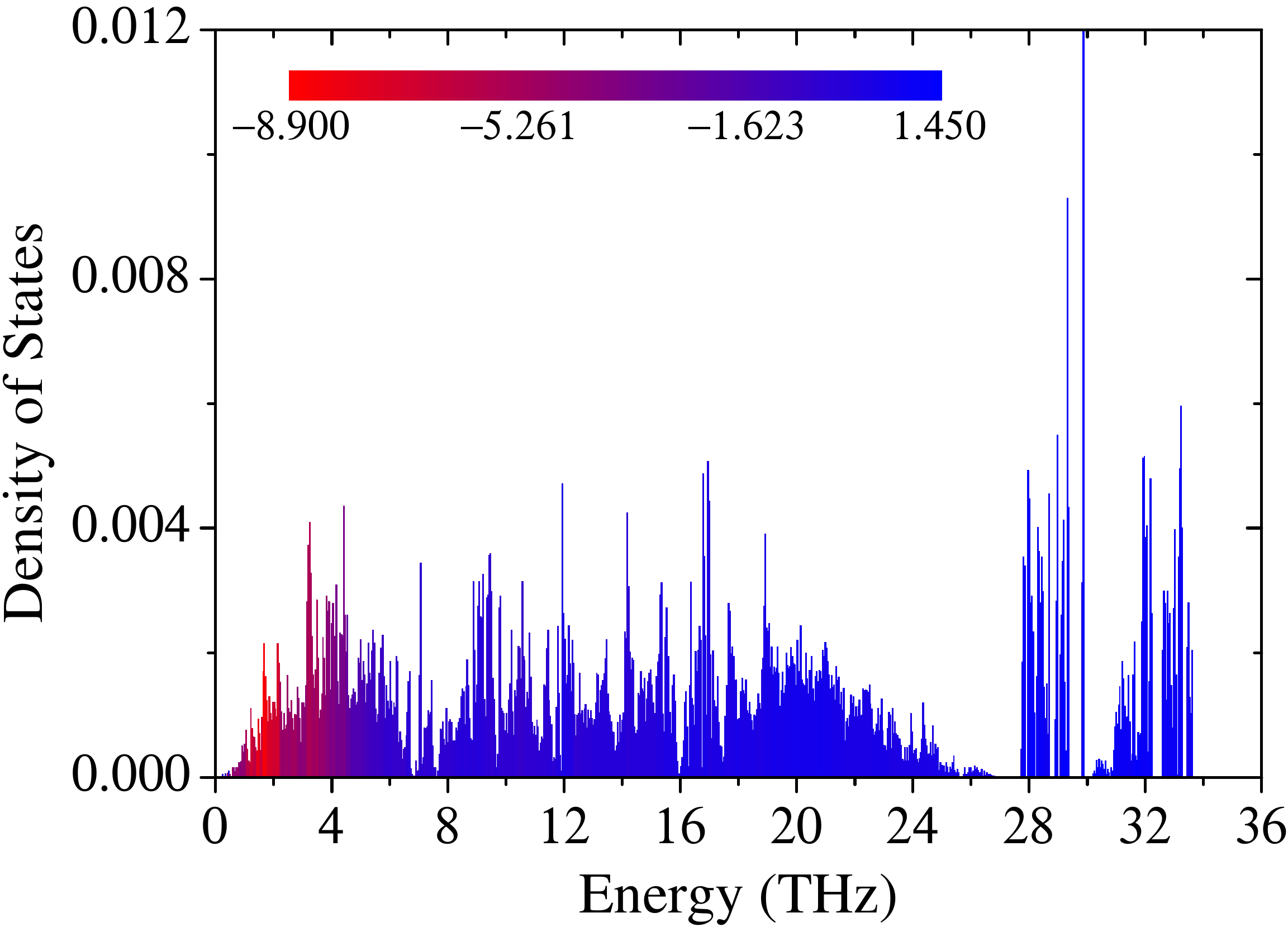}
\end{center}
\caption{\label{fig:FAU_dos} Calculated vibrational density of states of FAU using harmonic lattice dynamics, colored according to the values of Gr\"{u}neisen parameters (see text for details). Red highlights the most negative Gr\"{u}neisen parameters ($\leq -9$), while blue are for the positive values $\geq 1$.}
\end{figure}

In principle, at low temperature the values of $B_0$ and $B^{\prime}_0$ measured experimentally may differ from those obtained from a classical MD simulation due to quantum effects. However, we note that much of the important flexibility in NTE materials comes from vibrational modes having low frequencies, and these modes contribute the most to NTE through their large negative Gr\"{u}neisen parameters \cite{our_MD}. Taking the example of FAU, we have obtained the set of mode Gr\"{u}neisen parameters from the phonon frequencies calculated using expanded and contracted ($\pm$0.1$\%$) unit-cell volumes, and have colored the vibrational density of states according to the value of Gr\"{u}neisen parameter, as shown in figure~\ref{fig:FAU_dos} (plots of the other NTE zeolites are available in \cite{SI}). It is clear that modes with the most negative Gr\"{u}neisen parameters --- highlighted by red to light violet --- are around 1 THz (48 K) and span to the lowest frequency. This suggests that even at low temperature $\sim 50$ K, these modes will not be `frozen' out and can still be excited and contribute to NTE and pressure-induced softening of the material. In such a case, the classical MD results at low temperature would not have too much difference from the real quantum picture.

As shown in Table~\ref{results_table}, all the cubic zeolites that have NTE and are stable under compression show pressure-induced softening (negative $B^\prime$). It is interesting to note that BSV having almost zero thermal expansion shows almost zero $B^\prime$, and ANA with positive coefficient of thermal expansion shows positive $B^\prime$.

To show the consistency between the theory in Section II and the MD results, we calculate $B^{\prime}$ using Equation~\ref{eq18} and give the values of various terms in Table~\ref{table1}. The values of $\overline \gamma$ were obtained from the phonon frequencies calculated using expanded and contracted $\pm 0.05\%$ unit-cell volumes at zero pressure. The term $\partial \overline \gamma  /\partial p$ was calculated as $\partial \overline \gamma  /\partial p \approx \left( {\overline \gamma  _p  - \overline \gamma  _{p = 0} } \right)/p$ for a small $p$ around zero pressure. Combined with the values of $\tau/V_\mathrm{at}=N\tau/V$, one can see the validity of Equation~\ref{eq16}. The volume variation for the calculation of $\overline \gamma$ at each pressure is much smaller relative to the volume reduction at that pressure. $\partial ^2 \overline \gamma  /\partial p^2$ is calculated as $\partial ^2 \overline \gamma  /\partial p^2  \approx \left( {\overline \gamma  _{p_2 }  - 2\overline \gamma  _{p_1 }  + \overline \gamma  _{p = 0} } \right)/p_1 ^2$ for small pressures $p_2>p_1>0$. As we mentioned in Section II, since this term has a coefficient of $B$, it will contribute the most to the negative $B^{\prime}$ in Equation~\ref{eq18}.

The values in the last two columns of Table \ref{table1} show a reasonable degree of consistency. The main differences between the two is mainly due to the anharmonic processes that are missing in the calculations of $\overline \gamma$, $\partial \overline \gamma  /\partial p$ and $\partial ^2 \overline \gamma  /\partial p^2$ using harmonic lattice dynamics.

\begin{table*}[t]
\setlength{\tabcolsep}{12pt}
\caption{Calculated various terms in Equation~\ref{eq18} for all the cubic NTE zeolites, except for AST and MTN which only show NTE in their high-temperature phases. The second column of the table lists the values of $\tau/V_\mathrm{at}=N\tau/V$ at 300 K, with $\tau/V_\mathrm{at}$ the average volume per atom in the system. The third column lists the values of the overall Gr\"{u}neisen parameters, $\overline \gamma$. The sixth column lists the values of $B^{\prime}$ at zero temperature obtained from harmonic lattice dynamics. The seventh column lists the calculated $B^{\prime}$ using Equation~\ref{eq18}. The last column gives the values of $B^{\prime}$ from molecular dynamics at 300 K. }
\centering
\begin{tabular}{c| c| c| c| c| c| c| c}
\hline\hline
Zeolite & $\tau/V_\mathrm{at}$ (GPa) & $\overline \gamma$ & $\partial \overline \gamma  /\partial p$ (GPa$^{-1}$) & $\partial ^2 \overline \gamma  /\partial p^2$ (GPa$^{-2}$) & $B^{\prime}\left.\right|_{T=0}$ & \footnotemark[1]$B^{\prime}$ & \footnotemark[2]$B^{\prime}$ \\ [0.5ex]
  \hline
  FAU& $0.16$ &$-0.40$ &$-0.27$ &$-0.20$ &$2.4$ &$-2.7$ &$-2.6$ \\ [0.5ex]
  KFI& $0.18$ &$-0.62$ &$-0.35$ &$-0.18$ &$2.6$ &$-4.1$ &$-4.4$ \\ [0.5ex]
  RHO& $0.18$ &$-0.44$ &$-0.23$ &$-0.16$ &$-0.1$&$-5.4$ &$-4.0$ \\ [0.5ex]
  LTA& $0.18$ &$-0.85$ &$-0.60$ &$-0.90$ &$3.0$ &$-27.1$ &$-13.0$ \\ [0.5ex]
  MEP& $0.21$ &$-1.65$ &$-1.85$ &$-5.22$ &$-3.1$&$-192.1$&$-30.0$ \\ [0.5ex]
  TSC& $0.16$ &$-0.41$ &$-0.24$ &$-0.10$ &$0.5$ &$-2.2$ &$-4.9$ \\ [0.5ex]
  \hline
\end{tabular}
\footnotetext[1]{Calculated using Equation~\ref{eq18}.}
\footnotetext[2]{MD results at 300 K.}
\label{table1}
\end{table*}

\subsection{Correlation between the pressure-induced softening and NTE}


Negative values of $\overline \gamma$, $\partial \gamma /\partial p$ and $\partial ^2 \gamma /\partial p^2$  can be satisfied by assuming a pressure-induced strain ($e$) dependence for the frequency of the NTE modes, namely

\begin{eqnarray}\label{eq22}
\omega^2  = \omega _0^2  + Ce > 0,
\end{eqnarray}

\noindent where $\omega_0$ and $C$ are positive constants. Clearly, the frequency of the mode will decrease with more negative strain $e$, i.e. the mode will be softened on compression hence having negative Gr\"{u}neisen parameter

\begin{eqnarray}\label{eq23}
\gamma  =  - \frac{1}{2\omega^2 }\frac{{\partial \omega^2 }}{{\partial e}} =  - \frac{C}{{\omega _0^2  + Ce}} < 0
\end{eqnarray}

\noindent From Equation \ref{eq23}, we have

\begin{eqnarray}\label{eq24}
\frac{{\partial \gamma }}{{\partial p}} = \frac{{\partial e }}{{\partial p}} \frac{{\partial \gamma }}{{\partial e}} = - \frac{{C^2 }}{{B\left( {\omega _0^2  + Ce} \right)^2 }} < 0
\end{eqnarray}

\noindent and

\begin{eqnarray}\label{eq25}
\frac{{\partial ^2 \gamma }}{{\partial p^2 }} \approx \left(\frac{{\partial e }}{{\partial p}}\right)^2 \frac{{\partial ^2 \gamma }}{{\partial e^2}} = - \frac{{2C^3 }}{{B^2 \left( {\omega _0^2  + Ce} \right)^3 }} < 0
\end{eqnarray}

\noindent Thus, a material with whose mode frequency is reasonably represent by Equation~\ref{eq22} will show NTE with negative $\partial \gamma / \partial p$ as well as negative $\partial ^2 \gamma  /\partial p^2$, and therefore is likely to have negative $B^{\prime}$ on heating according to Equation~\ref{eq18}.

\begin{figure}[t]
\begin{center}
\includegraphics[width=8.5cm]{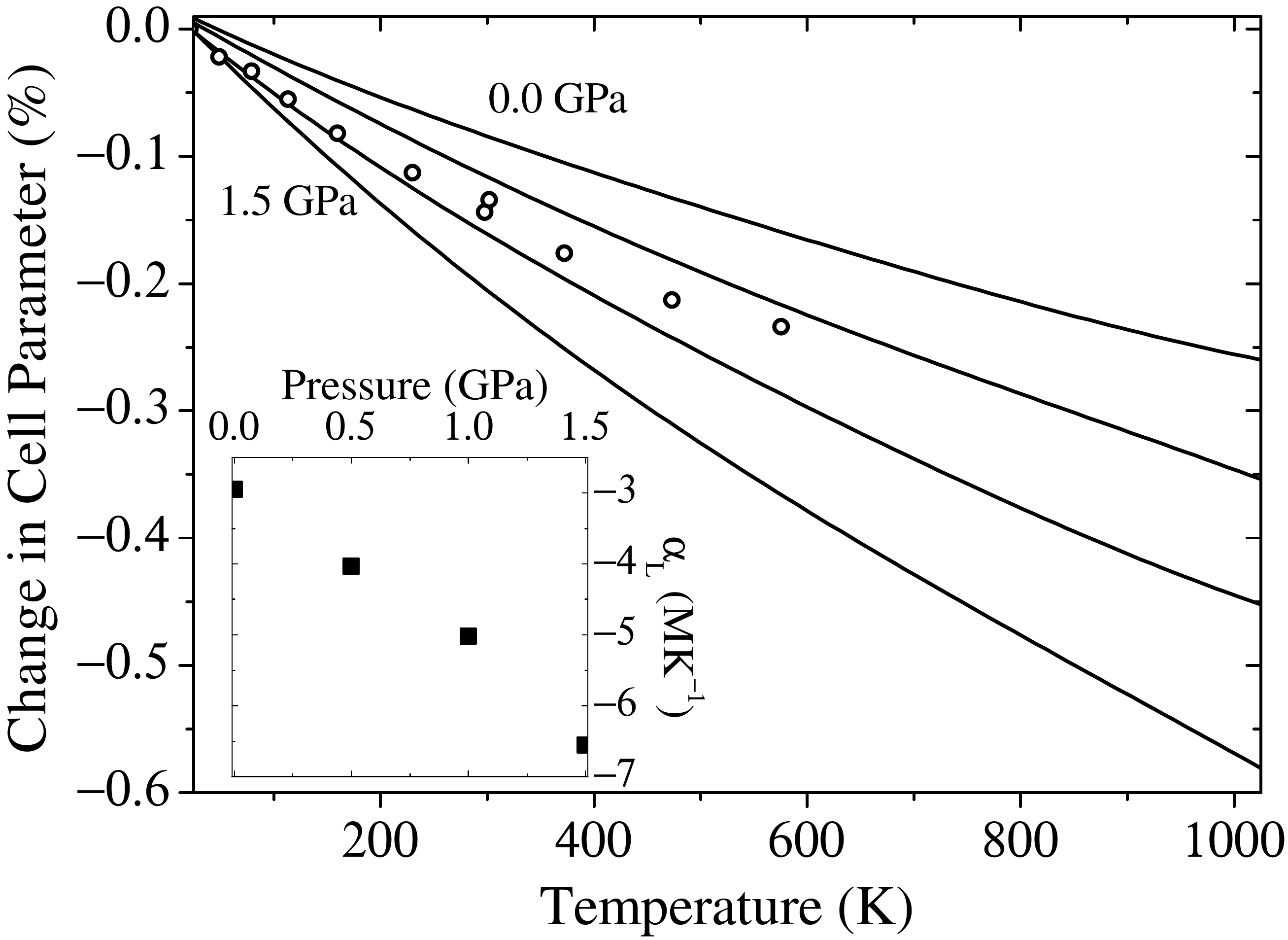}
\end{center}
\caption{\label{fig:FAUcell} Temperature dependence of the cell parameter of FAU at different pressures from MD simulations. The results agree well with the experimental data \cite{Attfield 1998}. The inset shows that the coefficient of thermal expansion ($\alpha$ at 300 K) of the material decreases on compression, i.e. the pressure-enhanced NTE.}
\end{figure}

One way in which pressure, elasticity and thermal fluctuations are linked is in the pressure-dependence of the coefficient of thermal expansion, $\alpha$, as shown by

\begin{eqnarray}\label{eq26}
\frac{1}{\alpha }\frac{{\partial \alpha }}{{\partial p}} = \frac{1}{{\overline \gamma  }}\frac{{\partial \overline \gamma  }}{{\partial p}} + \frac{1}{B}\left( {1 - B^{\prime}} \right)
\end{eqnarray}

\noindent For all the NTE zeolites (whose $\alpha <0$) listed in Table~\ref{table1} , both terms on the right-hand side of the equation would be positive, resulting in negative $\partial \alpha /\partial p$, i.e. the coefficient of thermal expansion becomes more negative under pressure. This pressure-enhancement of NTE has been confirmed by our MD results. For example, Figure~\ref{fig:FAUcell} shows both the simulation and the experimental data for FAU. Plots of this kind for the other zeolites are given in the supplemental material \cite{SI}.

\section{Conclusions}

The main conclusion from this study is that all the cubic zeolites that show NTE and are stable under pressure have negative $B^{\prime}$. This is a result from simulations and will need experimental verification, but the results are so overwhelmingly positive from simulation that we are confident they reflect the underlying physical processes. This lends strong support to our proposal that many NTE materials are likely to have pressure-induced softening.

The origin of the pressure-induced softening is rooted in the dependence of the frequencies of the NTE phonon modes on strain. With a simple form of frequency having positive dependence on pressure-induced strain, the phonon modes would not only have negative Gr\"{u}neisen parameters but also have negative first and second derivatives of the Gr\"{u}neisen parameter with respect to pressure, resulting in NTE as well as pressure-induced softening of the material.

With an increasing number of NTE materials being discovered, we suggest that there should be an increased focus on experimental searches for pressure-induced softening in these materials.

\begin{acknowledgments}
We gratefully acknowledge financial support from the CISS of Cambridge Overseas Trust (HF). MD simulations were performed using the CamGrid high-throughput environment of the University of Cambridge.
\end{acknowledgments}


\begin{thebibliography}{100}

\bibitem{Tsiok 1998}
O. B. Tsiok, V. V. Brazhkin, A. G. Lyapin, and L. G. Khvostantsev, {\it Phys. Rev. Lett.} {\bf 80}, 999 (1998).

\bibitem{Pantea 2006}
C. Pantea, A. Migliori, P. B. Littlewood, Y. Zhao, H. Ledbetter, J. C. Lashley, T. Kimura, J. Van Duijn, and G. R. Kowach, {\it Phys. Rev. B} {\bf 73}, 214118 (2006).

\bibitem{Chapman 2007}
K. W. Chapman and P. J. Chupas, {\it J. Am. Chem. Soc.} {\bf 129}, 10090 (2007).

\bibitem{Walker 2007}
A. M. Walker, L. A. Sullivan, K. Trachenko, R. P. Bruin, T. O. H. White, M. T. Dove, R. P. Tyer, I. T. Todorov, S. A. Wells, {\it J. Phys.: Condens. Matter} {\bf 19}, 275210 (2007).

\bibitem{Lind 2012}
C. Lind, {\it Materials} {\bf 5}, 1125 (2012).

\bibitem{Goodwin 2005}
A. L. Goodwin and C. J. Kepert, {\it Phys. Rev. B} {\bf 71}, R140301 (2005).

\bibitem{Goodwin2006}
A. L. Goodwin, {\it Phys. Rev. B} {\bf 74}, 134302 (2006).

\bibitem{our_MD}
H. Fang, M. T. Dove, L. H. N. Rimmer, and A. J. Misquitta {\it arXiv:1304.4789[cond-mat.mtrl-sci]} (2013).

\bibitem{our_Exp}
H. Fang, A. E. Phillips, M. T. Dove, M. G. Tucker, and A. L. Goodwin {\it arXiv:1306.1909[cond-mat.mtrl-sci]} (2013).

\bibitem{zeolite_note}
Detail information of these zeolites including the name and structure can be found at the on-line database: http://www.iza-structure.org/databases/.

\bibitem{Miller 2009}
W. Miller, C. W. Smith, D. S. Mackenzie, K. E. Evans, {\it J. Mater. Sci.} {\bf 44}, 5441 (2009).

\bibitem{Lightfoot 2001}
P. Lightfoot, D. A. Woodcock, M. J. Maple, L. A. Villaescusa, and P. A. Wright, {\it J. Mater. Chem.} {\bf 11}, 212 (2001).

\bibitem{Ashcroft 1976}
Neil W. Ashcroft and N. David Mermin, {\it Solid State Physics}, Brooks/Cole, a part of Cengage Learning, 1976.

\bibitem{Todorov 2006}
I. T. Todorov, W. Smith, K. Trachenko, and M. T. Dove, {\it J. Mater. Chem.} {\bf 16}, 1611 (2006).

\bibitem{Tsuneyuki 1988}
S. Tsuneyuki, M. Tsukada, H. Aoki, and Y. Matsui, {\it Phys. Rev. Lett.} {\bf 61}, 869 (1988).

\bibitem{William 1982}
William C. Swope, Hans C. Andersen, Peter H. Berens, and Kent R. Wilson, {\it J. Chem. Phys.} {\bf 76}, 637 (1982).

\bibitem{Hoover 1985}
W. G. Hoover, {\it Phys. Rev. A} {\bf 31}, 1695-1697 (1985).

\bibitem{Gale 1997}
J. D. Gale, {\it J. Chem. Soc., Faraday Trans.} {\bf 93}, 629 (1997).

\bibitem{Giddy 1993}
A. P. Giddy, M. T. Dove, G. S. Pawley, and V. Heine, {\it Acta Crystallographica A} {\bf 49}, 697703 (1993).

\bibitem{Hammonds 1994}
K. D. Hammonds, M. T. Dove, A. P. Giddy, and V.Heine, {\it American Mineralogist} {\bf 79}, 1207 (1994).

\bibitem{prepare}
Paper in preparation.

\bibitem{Attfield 1998}
M. P. Attfield and A. W. Sleight, {\it Chem. Commun.} {\bf 5}, 601 (1998).

\bibitem{Tschaufeser 1995}
P. Tschaufeser and S. C. Parker, {\it J. Phys. Chem.} {\bf 99}, 10609 (1995).

\bibitem{Park 1997}
S. H. Park, R. W. G. Kuntsleve, H. Graetsch, and H. Gies, {\it Stud. Surf. Science Catal.} {\bf 105}, 1989 (1997).

\bibitem{Colligan 2004}
M. Colligan, P. M. Forster, A. K. Cheetham, Y. Lee, T. Vogt, and J. A. Hriljac, {\it J. Am. Chem. Soc.} {\bf 126}, 12015 (2004).

\bibitem{Astala 2004}
R. Astala, S. M. Auerbach, and P. A. Monson, {\it J. Phys. Chem. B} {\bf 108}, 9208 (2004).

\bibitem{SI}
Supplemental Material.

\bibitem{EoSreview}
P. B. Roy and S. B. Roy, {\it J. Phys.: Condens. Matter} {\bf 17}, 6193 (2005).


\end{thebibliography}
\end{document}